# Mining Influential Spreaders in Complex Networks by an Effective Combination of the Degree and K-Shell

Shima Esfandiari
*Depatment of computer engineering*
*Shiraz University*
Shiraz, Iran
shima.esfandiari@hafez.shirazu.ac.ir

Seyed Mostafa Fakhrahmad
*Depatment of computer engineering*
*Shiraz University*
Shiraz, Iran
fakhrahmad@shirazu.ac.ir

*Abstract*— Graph mining is an important technique that used in many applications such as predicting and understanding behaviors and information dissemination within networks. One crucial aspect of graph mining is the identification and ranking of influential nodes, which has applications in various fields including marketing, social communications, and disease control. However, existing models and methods come with high computational complexity and may not accurately distinguish and identify influential nodes. This paper develops a method based on the k-shell index and degree centrality of nodes and their neighbors. Comparisons to previous works, such as Degree and Neighborhood information Centrality (DNC) and Neighborhood and Path Information Centrality (NPIC), are conducted. The evaluations, which include the correctness with Kendall's Tau, resolution with monotonicity index, correlation plots, and time complexity, demonstrate its superior results.

## I. INTRODUCTION

Graph mining allows us to extract valuable and actionable insights from graph data [1]. It enables us to predict and understand behaviors and information dissemination within networks [2]. This opens up possibilities for leveraging graph data in various domains such as marketing [3], advertising [4], and social communications [5] to enhance and optimize strategies. One crucial aspect of graph mining is the identification and ranking of influential nodes. In marketing, this involves finding people who have a great influence and can effectively carry out extensive advertising campaigns [6]. In social communications, it can be exemplified by preventing the spread of rumors or controlling information flow [7]. In the context of disease networks, it entails quarantining individuals who can effectively reduce or halt the spread of diseases [8]. These applications and many others [9,10] have motivated researchers to develop accurate and efficient methods for identifying influential nodes, providing valuable tools for decision-making and strategy optimization.

While models like the Susceptible-Infected-Recovered (SIR) model [11] can be employed for this purpose, they come with high computational complexity and may not be well-suited for real networks, which are typically complex [12]. Examining these complex networks contributes to a better understanding and deeper analysis of the real world. Notably, scale-free [13] and small-world [14] networks serve as examples.

The utilization of structural information from graphs has assisted researchers in offering alternative approaches to the SIR model. For instance degree centrality[15], closeness centrality (CC) [16], betweenness centrality (BC) [17], and k-shell index (KS) [18] have been employed as fundamental features. The Degree, as a local metric, exhibits low computational complexity but focuses solely on the lenght of node's neighborhood set. In contrast, the KS, BC, and CC represent global metrics. The k-shell stands out with its linear computational complexity [18], making it highly applicable in hybrid methodologies.

The gravity method [19] and its developments [20-23] employ the principles of Newton's law of gravity, making them hybrid approaches. In the gravity index, KS are treated as masse, while the distance is determined by the shortest path between two nodes. These two nodes consist of the target node and its r-level neighbors. However, the gravity centrality does not determine the optimal radius or level of neighbors. The Local Gravity Model (LGM) [20], on the other hand, finds this radius to be half of the network's diameter. Other methods, such as Laplacian Gravity Centrality (LGC) [21] and Effective Gravity Model (EGM) [22], utilize Laplacian centrality and effective distance, respectively.

Certain approaches focus on communities [24]. Clustering methods consider several critical factors, such as assigning higher values to nodes connected to multiple clusters or nodes within a robust cluster that has a high number of nodes and links [25,26]. These values are calculated using either entropy [26] or counting techniques [26]. Some distinguish between hub and bridge nodes by employing Newman's modularity [27].

Numerous approaches have been designed to identify influential nodes [15-29], each with its advantages and disadvantages. Some have high time complexity [16, 17, 21], while others exhibit low accuracy and resolution [15,18]. Additionally, certain methods require time-consuming parameter tuning [28,29]. This paper aims to develop a method that leverages the structural characteristics of the graph to offer superior accuracy and resolution compared to previous works. Moreover, this method should possess linear time complexity to efficiently handle large-scale real-world networks.

## II. RELATED WORK

This section introduces the SIR, baseline methods (DC and KS decomposition method), and Hybrid methods including Gravity, Local Gravity Model, Degree and Neighborhood information Centrality [28], and Neighborhood and path information centrality [29]. Finally, the limitation of each method is described.

### A. SIR model

It is used to assess the actual spreading ability of nodes. The SIR [11] categorizes nodes into three states: removed, infected, or susceptible. Initially, all nodes except the one that is infected are considered susceptible. At each step, any node in I state infect its neighbors of S with Beta probability and itself go to a R state with a probability represented as α (usually considered as 1). If a node enters the R state, it becomes immune to further infection. This process continues until no node changes state. The average percentage of nodes that have been enhanced after 100 iterations is considered as the actual influence ability of the node (which is the same node that was initially infected).

### B. Baseline methods

This part describes the degree and k-shell, which we further use in our evaluations.

- Degree centrality (DC) [15]: It quantifies the node's number of links or connections in a network.

- KS decomposition method [18]: It works by systematically identifying and assigning nodes to different shells or layers based on their connectivity within the network. Here is a step-by-step explanation of how it works:

1) Start with the original network.
2) Identify the nodes with the lowest degree (KS-index = 1), which are the nodes with only one connection.
3) Assign these nodes to the first shell (KS-index = 1).
4) Remove the nodes from the network.
5) Repeat steps 2-4 until all nodes have an assigned KS-index.
6) Move on to the next level of connectivity (KS-index = 2) and repeat steps 2-5.
7) Continue this process, incrementing the value of KS-index, until all nodes have an assigned KS.

In the end a hierarchical structure of shells is obtained. The nodes with the highest KS values, are considered to be more central or influential within the network, while nodes in the outer shells have lower connectivity and are considered less central.

### C. Hybrid methods

This part describes the hybrid methods that combine the degree or k-shell with other attributes, which we further use in our evaluations.

- Gravity [19]: The gravity formula is employed in this approach, utilizing the KS values as masses and the squared Euclidean distance as the distance metric. The calculation for node "i" is performed for its neighboring nodes. The paper specifically considers only three levels of neighbors in its analysis. It is defined as

$$Gravity(i) = \sum_{j \in \psi_{i,3}} \frac{ks_i * ks_j}{d_{ij}^2}$$

where $ks_i, ks_j, d_{ij}^2$ denote the KS of node i, j, and squared shortest path between i and j. The $\psi_{i,3}$ is the neighborhood set of node i that are at most in third level.

- Local Gravity Model (LGM) [20]: It finds the best level of neighbors for gravity formula in identifying influential nodes. It was the half of network average distance. The formula is

$$LGM(i) = \sum_{j \in \psi_{i,<d>/2}} \frac{ks_i * ks_j}{d_{ij}^2}$$

where $<d>$ represents the average distance.

- Degree and Neighborhood information Centrality (DNC) [28]: It utilizes a combination of the degree and the sum of local clustering coefficients of its direct neighbors. The parameter $\alpha$ is employed to adjust the degree proportionality and takes on a value greater than 0. In the paper, they set $\alpha$ to one. It is defined by

$$DNC(i) = k_i + \alpha \sum_{j \in \psi_{i,1}} C_j, \alpha > 0$$

where $C_j, k_i, \psi_{i,1}$ denote the local clustering coefficient, degree, and first-level neighbors of node i.

- Neighborhood and path information centrality (NPIC) [29]: This method incorporates two features: self-influence (SI) and path influence (PI). The SI combines the DC and the KS index. The PI considers the impact of all nodes on the target node and adjusts its value proportionally to the distance. The parameters $\alpha$ and $\beta$ are introduced to provide more flexibility in adjusting the impact of each feature and can be varied between 0.1 and 1. It is calculated by

$$SI(i) = \frac{(ks_i * k_i) + \alpha}{n}, PI(i) = \sum_{i \neq j} \frac{(ks_j * k_j) + \beta}{d_{ij}}$$

$$NPIC(i) = SI(i) * PI(i)$$

where $ks_i, k_i, n$ represents the KS-index, degree of node 'i' and the network size. The $d_{ij}$ shows the shortest path between 'i' and 'j'.

The degree centralty and KS are basic and simple approaches that are combined with other methods or enhanced in some way to identify influential nodes. The Gravity method has a combination of k-shell and distance, but the optimal levels of neighbors were not determined, which was suggested by LGM to be set as the half of network diameter. It also could be time-consuming for some networks. The NPIC and DNC are recent methods with free parameters which make them time-consuming. Moreover, the NPIC uses all nodes of the networks which makes it quadratic time complexity. Therefore, there is still a need for a method that has high correctness and low time complexity. In this paper, we proposed a method based on the KS index and the degree of its neighborhood to attain better correctness, time complexity, and resolution.

## III. Proposed method

To determine the diffusion capability of each node, it is important to consider both the node influence and the presence of powerful neighbors with a significant number of connections. To capture these aspects, we define two key features: Self-Influence (SI) and Neighborhood-Influence (NI). The Self-Influence feature quantifies the impact of the node itself, incorporating a combination of two structural properties of the graph: the DC and the KS-index. By leveraging the degree as a measure of local information and the KS-index as a measure of global information, which also accounts for linear time complexity, we can effectively capture both global and local aspects.

The Neighborhood-Influence considers two aspect for each neighbor: the node's influence that is calculated by its Self-Influence, and the feature belongs to its neighboring role that is the distance with target node. In other words, each node inherently has a self-influence equal to its SI value, the influence of more distant neighbors should decrease proportionally with their distance. To manage the time complexity and avoid incorporating irrelevant information, we limit the consideration of neighbors up to a certain distance, denoted as r. The value of r, which ranges from 1 to 3, is determined based on the specific characteristics of each network, as outlined in TABLE 1. By incorporating both Self-Influence and Neighborhood-Influence, we can effectively evaluate the influence capabilities of nodes. The SI and NI are calculated by

$$SI(i) = ks_i + k_i, \qquad NI(i) = \sum_{j \in \psi_{i,r}} \frac{SI(j)}{d_{ij}}$$

where $ks_i, k_i, \psi_{i,r}, d_{ij}$ represent the k-shell index, degree, r-level of neighborhood set for node 'i', and the distance between 'i' and 'j'. Finally, they combine with each other by

$$DKS(i) = SI(i) * NI(i) = (ks_i + k_i) * \sum_{j \in \psi_{i,r}} \frac{(ks_j + k_j)}{d_{ij}}$$

where DKS denotes the proposed method. The help for implementation is in Pseudo-code 1. The K-Shell index and Degree Centrality of all nodes are calculated in the first and second lines of the algorithm. Subsequently, lines 4 to 9 involve measuring the SI and NI for each node in the proposed method.

Pseudo-code 1. DKS (Input: radius, Graph (V, E) , Output: DKS of all nodes)

| | |
|---|---|
| c | KS <- Find K-shell value for all nodes of V |
| 2 | K <- Find Degree value for all nodes of V |
| 3 | DKS = [] |
| 4 | For all i in V: |
| 5 |     SI = KS[i]+K[i] |
| 6 |     NI=0 |
| 7 |     For all j in G.neighbors (i, level<=radius): |
| 8 |         NI += (KS[j]+K[j])/level |
| 9 |     DKS <- i, SI*NI |

## IV. Dataset & Evaluations

The Data set consists of four networks and the evaluation has four aspects correlation, correctness, resolution, and computational complexity.

### A. Dataset

The dataset consists of 4 complex networks of various sizes, with the smallest network, Dolphins, having 62 nodes and the largest network, Router, consisting of 5022 nodes. All these networks are included in the dataset [30]. More detailed information about the dataset can be seen in the TABLE 1. Note that the infection probability threshold calculated by

$$Beta_{th} = \frac{<k>}{<k^2> - <k>}$$

where $<k>$ and $<k^2>$ are the average degree and $2^{nd}$-order average degree. The radius is the parameter of DKS method.

TABLE 1. THE DATASET FUNDAMENTAL INFORMATION. THE N, M DENOTES THE NUMBER OF NODES AND EDGES. THE <D> REPRESENTS THE AVERAGE DISTANCE THE R AND BETA$_{TH}$ ARE DEGREE ASSORTATIVITY AND THE INFECTION PROBABILITY THRESHOLD. THE RADIUS IS OUR PARAMETER IN DKS METHOD. THE <K> DENOTES THE AVGERAGE DEGREE

| Network | n | m | <k> | <d> | r | Beta$_{th}$ | radius |
|---|---|---|---|---|---|---|---|
| Dolphins | 62 | 159 | 5.1 | 3.357 | -0.044 | 0.16 | 3 |
| Netsci | 379 | 914 | 4.8 | 6.042 | -0.082 | 0.14 | 3 |
| Power | 4941 | 6594 | 2.7 | 18.989 | 0.003 | 0.26 | 2 |
| Router | 5022 | 6258 | 2.5 | 6.449 | -0.138 | 0.08 | 3 |

### B. Correlation

In this section, we focus on the correlation of the DKS and any some of benchmark methods. In Fig. 1, Fig. 2, in addition to the correlation, we have also utilized the third dimension, represented by color, which signifies the spreading ability of each node. The methods are Degree, K-Shell, Local Gravity Model, DNC, and NPIC. They are applied to dataset networks. As observed, the proposed method exhibits the highest similarity to the LGM.

The figures presented in Fig. 1, Fig. 2 offer a comprehensive overview of the performance of each method in comparison to the proposed approach. For instance, the k-shell and degree methods exhibit comparatively weaker performance. On the other hand, the DKS has higher correctness in the Power when compared to the DNC method. To ensure a more accurate evaluation, metrics such as Kendall's Tau [31] for correctness and the imprecision function [32] for resolution are employed.

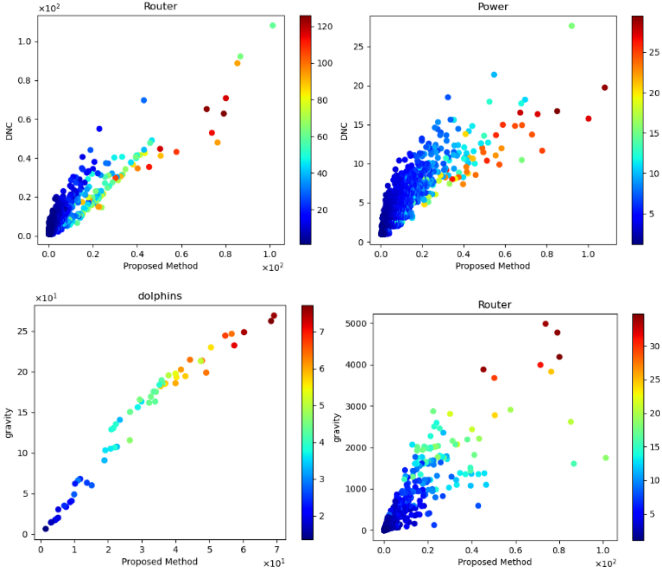

Fig. 1. The Correlation between the DKS and other recent methods. The colors indicate the spreading ability calculated by the SIR model.

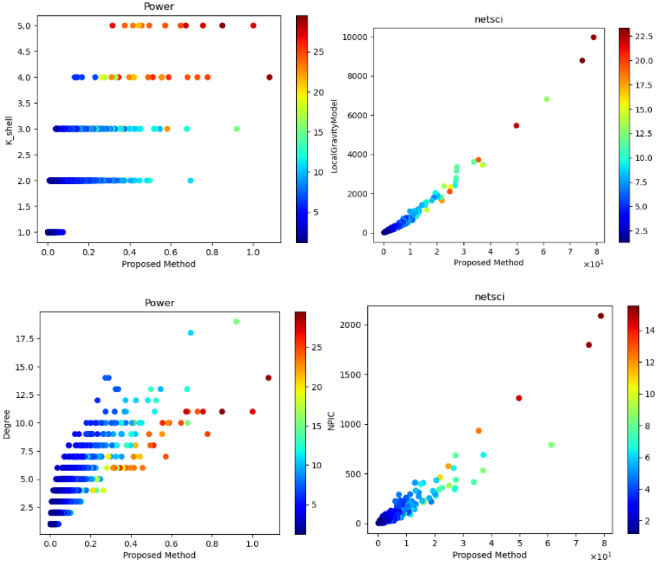

Fig. 2. The Correlation between the DKS and other recent methods. The colors indicate the real spreading ability.

### C. Kendall's Tau-b (Correctness)

It [31] is a measure utilized to assess the correlation of 2 ranked variables. In this context, the ranked variables pertain to the ordering of nodes based on the SIR and a novel method. It ranges in (-1,1), with higher values indicating a stronger and more accurate relationship between the two methods. It is defined by

$$Tau(Rank\ List_1, Rank\ List_2) = \frac{n_{con} - n_{dis}}{\sqrt{(N_{pairs} - n_{t1}) \times (N_{pairs} - n_{t2})}}$$

Where $N_{pairs}$, $n_{dis}$, $n_{con}$ denote the total number of pairs, discordant pairs, and the number of concordant pairs (pairs where the ranks of both variables have the same direction), respectively. Meanwhile, $n_{t1}$ and $n_{t2}$ indicate the counts of tied ranks in the first and second rank lists, respectively.

Fig. 3 and Fig. 4 illustrate the comparison of the correctness of the DKS method with six recent methods in the networks of the dataset. The horizontal axis shows the spreading probabilities and the vertical axis represents the value of tau.

In all networks, the proposed method consistently exhibits higher values of Tau, particularly in the vicinity of the infection probability threshold (As mentioned in TABLE 1). In the Dolphins and Netsci, the DKS ranks first in terms of correctness for both the threshold probability and higher infection probabilities. In the Power, the proposed method outperforms other methods for infection probabilities ranging from 0.13 to 0.28 where beta$_{th}$=0.26. In the Router, the values of the DKS and Gravity methods are very close to each other but the DKS slightly better.

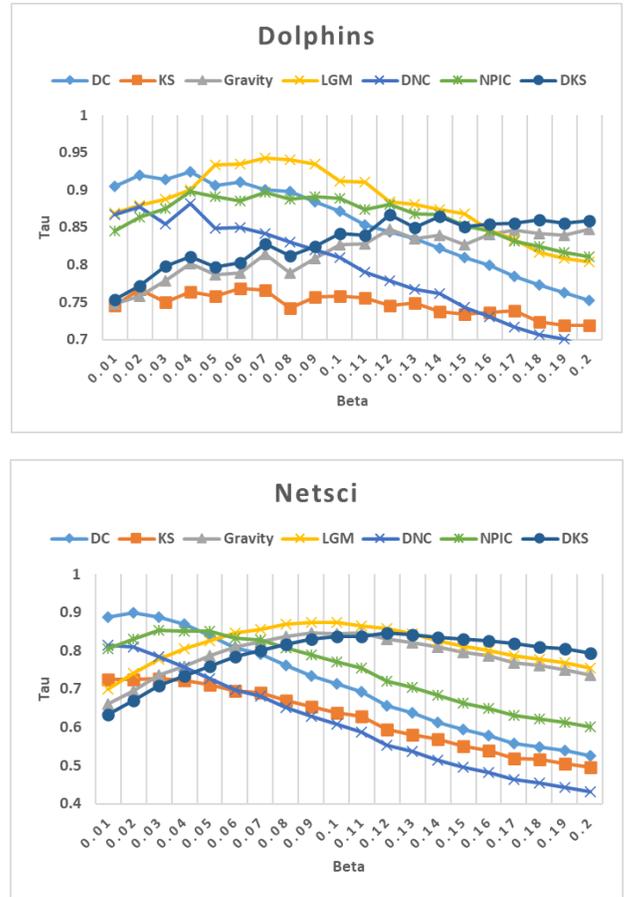

Fig. 3. The correctness of 7 methods in the Dolphins and Netsci

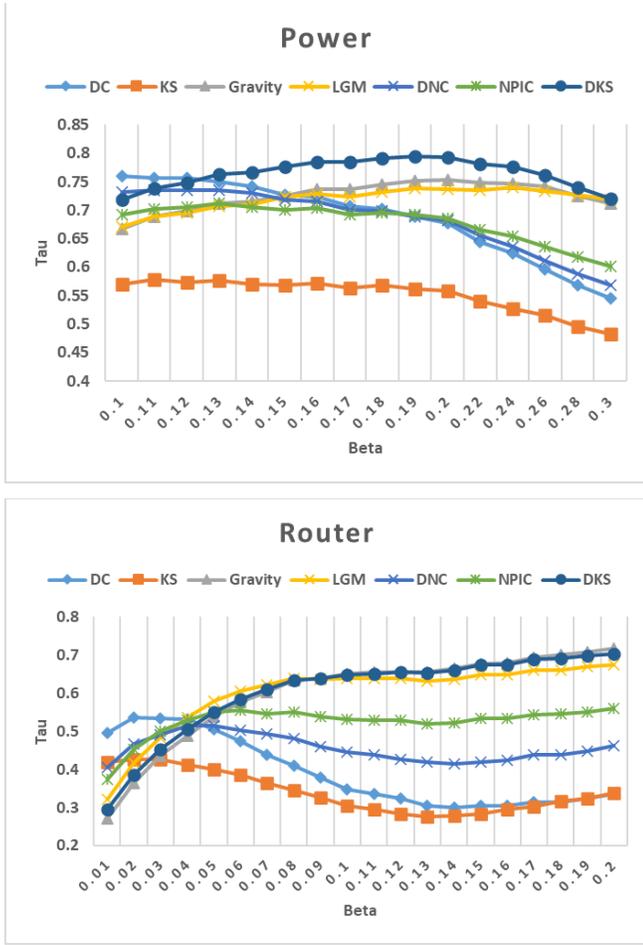

Fig. 4. The correctness of 7 methods in the Power and Router

## D. Monotonicity index (Resolution)

The monotonicity index is employed as a means of assessing the resolution of a given methodology in the identification of influential nodes [32]. This index assumes values between 0 and 1, where a value of 1 represents the optimal outcome in terms of successfully distinguishing between nodes. The formula utilized to calculate the monotonicity index is

$$M(RM) = [1 - \frac{\sum_{r \in RM} N_r(N_r - 1)}{N(N-1)}]^2$$

where $N$ is the network's size, RM is to the ranking method employed, while $N_r$ is the number of nodes with 'r' rank.

TABLE 2 shows the values of the monotonicity index for the 6 methods and the DKS in the dataset networks. The bold number shows the highest value in that network. It highlights the effectiveness of Gravity, NPCI, and DKS in accurately differentiating between the influence capabilities of different nodes. The DKS addresses the limitation of the DC and KS-index methods, which assume equal influence for a large number of nodes. It is happened by intelligently combining these two features, we have successfully overcome this challenge in the proposed method.

TABLE 2. THE MONOTONICITY INDEX OF EIGHT METHODS IN NETWORKS

| Network | DC | KS | Gravity | LGM | DNC | NPCI | DKS |
|---|---|---|---|---|---|---|---|
| Dolphins | 0.83 | 0.38 | **1.00** | 0.98 | **1.00** | **1.00** | **1.00** |
| Netsci | 0.76 | 0.64 | **1.00** | **1.00** | 0.99 | **1.00** | **1.00** |
| Power | 0.59 | 0.25 | **1.00** | **1.00** | 0.81 | **1.00** | **1.00** |
| Router | 0.29 | 0.07 | **1.00** | **1.00** | 0.07 | **1.00** | **1.00** |

## E. Computational complexity

Evaluations for identifying influential nodes are an essential component because any proposed method must apply to real networks, which typically have large sizes.

The proposed method includes three features: distance, DC, and KS-index. The time complexity of the KS and DC are equals to the number of edges (m) and nodes (n), respectively. Calculating the SI is equals to O(m) and NI for r <3 is in average O(m*2). In general, the DKS time complexity of the is O(2*m) which shows that the DKS is a linear method.

## V. CONCLUSION

Graph mining has emerged as a widely applicable field in today's world, with applications ranging from social networks to business and rumor spreading. This paper focuses on one of its subsets that find and rank the ability of nodes in term of spreading ability. The proposed method combines local and global graph features, including DC, KS-index, and distance, to form the DKS method. Evaluations from four perspectives demonstrate the superiority of the DKS method over six previous approaches. The real spreading ability of nodes was simulated using the SIR model in 4 real networks and compared with the proposed method and previous methods. Due to its linear nature, the proposed method is also executable in large-scale networks.